\documentclass[aps,prl,secnumarabic,
twocolumn,
groupedaddress,showpacs,amsmath,amssymb,superscriptaddress]{revtex4}

\usepackage{graphicx}
\usepackage{bm}

\begin{document}

\title{New Approach to Metal--Insulator Phase Transition Kinetics
in Magnetic Field.}

\author{L.B. Dubovskii}
\email[]{Dubovskiy_LB@rrcki} \affiliation{Theoretical division , NRC"Kurchatov Institute", 123182 Moscow, Russia}
\affiliation{chair of theoretical physics, MFTI,  141700 Moscow region}

\begin{abstract}
The metal--insulator phase transition is considered on the basis of Ginzburg-Landau type
equations with two different order parameters.
An inclusion of magnetic field in this picture is an important step for understanding of behavior of the metal--insulator phase transition kinetics. The magnetic field leads to various singularities of the surface tension and results in drastic variations of the phase transition kinetics. The strongest singularity is due to Landau diamagnetism and determines anomalous features of MI transition kinetics. This singularity supports the well known experimental fact that almost all semimetals behave like diamagnetic materials.
\end{abstract}


\maketitle
We investigate phase transitions from an insulator phase to a metal phase (MI phase transitions).
Any MI transition in a crystalline material, at any rate at zero temperature, must be a transition
from a situation in which electronic bands do not overlap to a situation when they do
\cite{NFMott}.
One of the overlapped bands is initially empty and gets carriers of electron type. The other one
is initially entirely filled with electrons and gets carriers of hole type due to overlapping of the
bands. Small band-crossing leads to two-band metallic state with small numbers of electrons and holes
carriers per atomic cell. In addition, the numbers of holes and electrons
are equal to each other. The situation is quite typical for common semimetals.
\par
The MI phase transitions are the phase transitions of the first order. The description of these
transitions is usually based on the common approach to the first order phase transition theory
\cite{LandLif}. The leading parameter which provides the MI phase transition is the material density $\rho(\vec r)$
of the phases. This parameter determines the first order type of the phase transition at a critical value of $\rho _c$. Here $\rho _c$ is  a function of pressure $p$ and temperature $T$ ($\rho _c =\rho _c (p,T)$). Generally, the pressure $p$ is the leading parameter which governs the MI transition. A growth of density
of the insulator phase leads to overlapping of two electronic bands.
One of these bands is empty and the other one is filled with electrons. By overlapping of such bands
the MI phase transition occurs with an appearance of a new order parameter which characterizes the
density of current carriers in the system. This parameter equals identically zero in the insulator
phase and grows from zero at the MI interface. This parameter exhibits itself as the second order phase transition parameter.
\par
So, we can describe the MI phase transition by two order parameters (cf. \cite{Lubensky, JinwuYe, dub, dub1}).
One of the order parameters is the material density distribution $\rho (\vec r)$ either of metal phase or of insulating one.
The second parameter is the density of electrons and holes in two bands. The density of electrons and holes
 can be described by a vector in the Hilbert space with electronic and holes components $\Upsilon ^{+} (\vec r)=
( u_{\uparrow} ^* (\vec r), u_ {\downarrow } ^* (\vec r), v_ {\uparrow} ^* (\vec r), v_ {\downarrow} ^* (\vec r) )$.
The components of the vector $\Upsilon (\vec r)$ are functions of electrons for up and down spins
$(u_{\uparrow}(\vec r), u_{\downarrow}(\vec r))$ and functions of holes for up and down spins
$(v_{\uparrow}(\vec r), v_{\downarrow}(\vec r))$. In the absence of magnetic field all these components of the vector
$\Upsilon (\vec r)$ are real functions. We can put the following invariant scalar quantity
\begin{equation}\label{glr2}
n(\vec r) =n_{e}(\vec r)+n_{h}(\vec r)\equiv\left({\Upsilon ^{+}\Upsilon }\right)\, ,
\end{equation}
$n_{e}(\vec r)=|u_{\uparrow}(\vec r)|^2 +|u_{\downarrow}(\vec r)|^2\, , \,\,\,
n_{h}(\vec r)=|v_{\uparrow}(\vec r)|^2 +|v_{\downarrow}(\vec r)|^2\, .$
It determines densities of electrons and holes in the metal phase and becomes identically zero
in the insulating phase at $T=0$. We take the identity $n_{h}(\vec r)=n_{e}(\vec r)$ due to the
local electrical neutrality within the metal phase.
\par
In the vicinity of MI phase transition, the Ginzburg-Landau (GL) functional can be assumed as a sum of contributions
from order parameters $\rho(\vec r)$, $\Upsilon (\vec r)$: 
\begin{eqnarray}\label{glrn1}
\Phi=\Phi\{\rho ,\Upsilon , \Upsilon ^{+} \}=\int d\vec{r}\emph{F}(\rho , \Upsilon , \Upsilon ^{+} )\, , \nonumber
\\\emph{F}=\emph{F}_{\rho}(\rho (\vec r))+\emph{F}_{\Upsilon}(\Upsilon (\vec r))+
\emph{F}_{int}(\rho (\vec r),\Upsilon (\vec r))   .
\end{eqnarray}

The parameter $\rho(\vec r)$ describes the first order phase transition between insulating and
metal phases whereas the vector function $\Upsilon (\vec r)$ of electrons and holes describes
the second order phase transition.
We expand the functional
$\emph{F}_{\rho}(\rho (\vec r))$ according to Cahn -- Hillard approximation \cite{CahnHillard},
see also \cite{LifKagan}:
\begin{equation}\label{glrnn1}
\emph{F}_{\rho}(\rho (\vec r))=\varphi (\rho ) +\frac{1}{2}\lambda (\rho )(\nabla\rho )^{2}
\end{equation}
We use the GL expansion \cite{LandLif2} for the vector function
$\Upsilon (\vec r)$
$$\emph{F}_{\Upsilon}(\Upsilon (\vec r))=\alpha\left({\Upsilon ^{+}\Upsilon}\right)+
\frac{1}{2}\beta{\left({\Upsilon ^{+}\Upsilon }\right)}^{2}+\frac{\hbar ^2}{2m}
\left(\nabla \Upsilon ^{+}(\vec r)\nabla \Upsilon (\vec r)\right) ;$$
\begin{equation}\label{glrnn2}
\alpha =\aleph (\rho-\rho _{c})
\end{equation}
Here we have expanded $\alpha$ linearly over $\rho$ in the vicinity of $\rho _{c}$. It is in analogy with the well known expansion of the appropriate parameter in the Landau theory of the second order phase transition on the temperature $T$ \cite{LandLif}. The proposed expansion is convenient for an investigation of the MI phase transitions at a fixed temperature, say at $T=0$.
For simplicity, we have taken
the masses of holes and electrons to be equal to each other.
Also the interaction term $\emph{F}_{int}(\rho (\vec r),\Upsilon (\vec r))$ is taken as a linear
function of $n(\vec r)$ (cf. \cite{Lubensky,JinwuYe, dub}):
\begin{equation}\label{glrnn3}
\emph{F}_{int}(\rho (\vec r),\Upsilon (\vec r))=g(\rho)\left({\Upsilon ^{+}\Upsilon }\right)
\end{equation}
The variation of $\Phi=\Phi\{\rho ,\Upsilon , \Upsilon ^{+} \}$ over $\rho (\vec r)$ and $\Upsilon ^{+}(\vec r)$
gives the self consistent equations of GL type for MI phase transition:
\begin{eqnarray}\label{gln8}
\lambda ^{1/2}(\rho)\nabla\left(\lambda ^{1/2}(\rho)\nabla\rho (\vec r)\right)=\frac{d\varphi}{d\rho}(\rho)+
\frac{dg}{d\rho}\left({\Upsilon ^{+}\Upsilon }\right)\nonumber
\\
-\frac{\hbar ^2}{2m}\nabla ^2 \Upsilon
+\left(\alpha +g\right)\Upsilon+\beta \left({\Upsilon ^{+}\Upsilon }\right) \Upsilon=0\,\,\,\,\,\,\,\,
\end{eqnarray}
In the case $g\equiv 0$,  the parameters $\rho $ and $\Upsilon $ are independent and the surface tension $\Sigma $ at the interface between the metallic and the insulating phases can be calculated separately
for  $\rho$ ($\Sigma _{\rho}$) and  $\Upsilon $ ($\Sigma _{\Upsilon}$).
 The surface tension $\Sigma _{\rho}$  can be calculated straightforward
 (\cite{LifKagan, dub}):
\begin{equation}\label{glq}
\Sigma _{\rho} =\int _{\rho _{ins}} ^{\rho _{met}} d\rho\sqrt{\varphi (\rho)\lambda (\rho )/2\, }
\end{equation}
It should be emphasized that the
value of  $\Sigma _{\rho}$  is of the order of the value of surface tension at the interface between an
insulator and vacuum. It is well known from experiments \cite{kik} that the characteristic values
of the surface tension at the interface between an insulator and vacuum are at least an order of magnitude
lower than those at the interface between a metal and vacuum. The physical mechanism underlying this
difference is related to the presence of conduction electrons penetrating from the interface to vacuum at a
distance of the order of the separation between atomic layers in the metal \cite{Partenskiy}. So,
\begin{equation}\nonumber
\Sigma _{\Upsilon}\gg\Sigma _{\rho}.
\end{equation}
In most cases we can neglect $\Sigma _{\rho}$ in comparison with $\Sigma _{\Upsilon}$.
\par
Now we consider the behavior of the system in magnetic field.
The investigation of MI interface in the presence of magnetic field is a remarkably delicate task.
The problem is quite similar for introducing of a magnetic field in Kohn-Sham scheme  \cite{VRG}
based on \cite{VR}.
One can understand the basic difficulty by considering the expression for the orbital current density:
\begin{equation}\nonumber
\vec j(\vec r)=\vec j_p (\vec r)+\frac{e}{mc}n(\vec r)\vec A(\vec r)
\end{equation}
For short, we assume the situation with only one conductivity band. The first term $\vec j_p (\vec r)$ is
referred to as "paramagnetic current density,"  and measures the density of canonical momentum in the
wave function. It depends explicitly on the electron density $n(\vec r)$ and the
vector potential $\vec A(\vec r)$. Here we should emphasize that the situation in normal metals differs
drastically from that in superconductors. In normal metals the electron density $n(\vec r)$ is a
function of coordinates $\vec r$ whereas in the superconductors the electron density $n(\vec r)$ is an integer
number of electron charge value \cite{GL} and does not depend on coordinates $\vec r$ at all.
 The integer value is just $2e$ according to the microscopic theory of superconductivity \cite{dG}.
 The brilliant discussion of the problem in superconductors can be found in \cite{MEG}. In \cite{VRG} it is
  assumed that the fields, state and its energy are uniquely determined by the densities,
and  the paramagnetic current density. This statement would have not been true if we had attempted to work
with the physical current densities $\vec j(\vec r)$ as basic variable. In that case one would still have
the freedom of operating a gauge transformation on the vector potential, and hence multiplying the
wave function by a coordinate-dependent phase factor, \textit{without changing} the physical current
density. Therefore, the state wave function is \textit{not} a unique function of $\vec j(\vec r) $.
 Working with the paramagnetic current density $\vec j_p$ eliminates this ambiguity.
In the presence of a magnetic field $\vec B$ the mapping between ground-state energy and ground-state
wave function is not one to one. To recover a one to one mapping it is necessary to supplement the number
density distribution by spin- and current- densities which play the same role with respect to an external field
$\vec B$, and vector potential $\vec A$, as the density does with respect to ordinary scalar potential.
In summary, the vector potential $\vec A$ should not be involved in the potential $\Phi$ for normal metals because the gauge invariance is violated.
\par
We assume (cf. \cite{RC,GH}) that the state and its wave function are determined uniquely by $\vec B(\vec r)$
and $\rho (\vec r)$. The inclusion of paramagnetic effects can
be carried out straightforward by adding an additional term into the functional (\ref{glrnn2}):
\begin{equation}\label{glrnn2phn}
\emph{F}_{para}(\vec {B} ,\Upsilon , \Upsilon ^{+} )=\mu _{ph}\left(\Upsilon ^{+}
\vec B\vec{\Sigma}\Upsilon\right),
\vec{\Sigma}= \begin{pmatrix} \vec{\tau _u}\,\,\,\,\,\,\,\,\,\, 0\\ 0\,\,\,\,\,\,\,\,\,\,\vec{\tau _v}
\end{pmatrix}
\end{equation}
$\mu _{ph}$ is a phenomenological constant,  $\vec{\tau _u} = \vec{\tau _v}$ are Pauli spinors \cite{RC}
correspondingly for electrons and holes.
\par
The functional $\Phi\{\rho ,\Upsilon , \Upsilon ^{+} \}$ (\ref{glrn1}) with the paramagnetic
term (\ref{glrnn2phn}) gives
the energy spectrum  $\varepsilon _k$ of the system and the thermodynamical potential
$\Omega (T,V,\mu)$ \cite{LandLif}:
\begin{equation}\nonumber
\Omega =-T\sum _k \ln \left(1+\exp \frac{\mu -\varepsilon _k}{T}\right)\,  ,\,\,\,
\vec M(\vec r)=-\frac{\partial \Omega}{\partial \vec B(\vec r)}.
\end{equation}
$\vec M(\vec r)$ is the magnetic moment of the metal and we assume that it  depends exclusively on the electron density $n(\vec r)$ or
$\left({\Upsilon ^{+}\Upsilon }\right)$ (see (\ref{glr2})) and does not depend on $\rho (\vec r)$, i.e.
\begin{equation}\label{glpinh1mn}
\vec M(\vec r)=\vec M(n(\vec r))\,\, ,\,\,\, n(\vec r)=\Upsilon ^{+} (\vec r)\Upsilon (\vec r).
\end{equation}
 The magnetic field
can be introduced as follows: $\,\,\,\,\,\,\,\,\,\,\,\,\,\,\,\,\,\,\,\,\,\,\,\,\,\,\,\,\,\,\,
\vec H(\vec r)=\vec B(\vec r)-4\pi\vec M(\vec r).$
\par
The thermodynamical potential $\Phi\{\rho ,\Upsilon , \Upsilon ^{+} \}$ (\ref{glrn1}) with the additional
 term (\ref{glrnn2phn}) is a function of the independent variables $T, V, \vec{B}$.
It is inconvenient since $\vec{B}$ means the average of the microscopic magnetic field in the specimen.
The value $\vec{B}$ cannot be fixed experimentally. Therefore we should fix the external magnetic field
$\vec H$. It can  be done by adding (see \cite{LLe, AAA}) the following value:
$$-\frac{{\vec H}{\vec B}}{4\pi}+\frac{{\vec H}^2}{8\pi}\equiv -\vec H\vec M-\frac{{\vec H}^2}{8\pi} $$
 and consider the other thermodynamic potential with the independent variable
$\vec{H}$:
\begin{eqnarray}\label{glrn1n}
\Phi _H\{\rho ,\Upsilon , \Upsilon ^{+} \}=\int d\vec{r}\emph{F}_H(\rho , \Upsilon , \Upsilon ^{+} )\nonumber
\\
\emph{F}_H(\rho , \Upsilon , \Upsilon ^{+} )=\emph{F}_{\rho}(\rho (\vec r))+\digamma _{\Upsilon,\vec H}+
\emph{F}_{int}(\rho (\vec r),\Upsilon (\vec r))
\end{eqnarray}
\begin{equation}\nonumber
\digamma _{\Upsilon,\vec H}=\emph{F}_{\Upsilon}(\Upsilon (\vec r))
+\emph{F}_{para}(\vec {H}+
4\pi\vec{M} ,\Upsilon , \Upsilon ^{+})-\vec H\vec M -\frac{{\vec H}^2}{8\pi}.
\end{equation}
The variation of $\Phi=\Phi\{\rho ,\Upsilon , \Upsilon ^{+} \}$ over $\Upsilon ^{+}(\vec r)$ for
the fixed independent variable $\vec{H}$ gives the equation
for $\Upsilon$:
\begin{equation}\label{glpinh3nm}
-\frac{\hbar ^2}{2m}\left(\frac{d^2 \Upsilon}{dx^2}\right)
+\left(\alpha +g\rho \right)\Upsilon+\beta \left({\Upsilon ^{+}\Upsilon}\right) \Upsilon
+\mathbb{H}\Upsilon=0\, ;
\end{equation}
\begin{equation}\nonumber
\mathbb{H}\Upsilon =\mu _{ph}\begin{pmatrix}H_z  & H_x  &0&0\\ H_x
 & -H_z  &0&0\\0&0&H_z & H_x \\0&0&H_x & -
 H_z
\end{pmatrix}\Upsilon -\vec H\frac{\partial\vec M}{\partial ( \Upsilon ^{+}\Upsilon )}\Upsilon  \, .
\end{equation}
Now we take $g\equiv0$  and calculate the surface tension $\Sigma _{\Upsilon}$. One by one, we shall consider two cases: 1) $\vec M \equiv 0 $ and
2) $ \mu _{ph}\equiv 0$. In the first case
 components of $\Upsilon (x)$ can be taken as real functions and an exact solution can be found.
The first two matrix rows of Eq. (\ref{glpinh3nm})
are not connected with the other two rows and they can be rewritten as the following set of two coupled
equations for the components $\Upsilon$ :
\begin{eqnarray}\label{glh3x}
\hat{L}(u_\uparrow , u_\downarrow ) u_{\uparrow}
+\mu _{ph} (H_xu_{\downarrow} + H_zu_{\uparrow})=0\nonumber
\\
\hat{L}(u_\uparrow , u_\downarrow ) u_{\downarrow}
+\mu _{ph} (H_xu_{\uparrow} - H_zu_{\downarrow})=0\, .
\end{eqnarray}
Here $\hat{L}(u, v)$ is the following linear differential operator:
\begin{equation}\nonumber
\hat{L}(u, v)\varphi=-\frac{\hbar ^2}{2m}\frac{d^2 \varphi}{dx^2}+
\alpha \varphi +\beta \left(|u|^2+|v|^2\right)\varphi\, .
\end{equation}
Solution of the system (\ref{glh3x}) can be sought in the form:
\begin{equation}\nonumber
u_{\downarrow}(x)=qu_{\uparrow}(x)\,\, ,
\end{equation}
where $q$ depends on $H_x$ and $H_z$.
The both equations of the system $(\ref{glh3x})$ become identical to each
other under the following condition:
$qH_x + H_z={H_x}/{q} - H_z\, .$
\par
So, we have
 $q=q_{\pm}$,
\begin{equation}\nonumber
 q_{+}=\frac{|H_x|}{\sqrt{H_x ^2 +H_z ^2}+H_z sgn{H_x}}\,\,\,\,\, ,\,\,   \,\,\,\,\,
 q_{-}=-\frac{1}{q_{+}}\, .
\end{equation}
The first equation of the system $(\ref{glh3x})$ can be rewritten:
\begin{eqnarray}\label{glh5xym}
-\frac{\hbar ^2}{2m}\frac{d^2 u_{\uparrow}}{dx^2}+\tilde{\alpha}u_{\uparrow}+\tilde{\beta}(1+q^2)
 {|u_{\uparrow}|}^2 u_{\uparrow} =0\,\, ;\,\,\,\,\nonumber
\\
\tilde{\alpha}=\alpha+\mu _p (q H_x  + H_z )\, ,\,\,\tilde{\beta}=\beta  (1+q^2)\, ;\,\,
q=q_{\pm}\,\, .
\end{eqnarray}
We can also $\tilde\alpha$ represent as follows
$$\tilde{\alpha}=a(\rho -\rho _{c \vec H})\, ,\,\,\,\,\,\,\,
\rho _{c \vec H}=\rho _{c 0}-\mu _{ph} Hsgn{H_x}a^{-1}\,\, ,\,\, $$
$H=\sqrt{H_x ^2 +H_z ^2}\, .$
\par
Equation (\ref{glh5xym}) has two solutions: (i) a homogeneous solution $u_{\uparrow}\equiv 0$
corresponding to the insulating state and (ii) the solution $(u_{\uparrow}^0 , u_{\downarrow}^0)$ having the absolute values
 \begin{equation}\label{glh3xx}
 |u_{\uparrow} ^0 | ^2 =-\frac{\tilde{\alpha}}{\tilde{\beta}(1+q^2)}\, ;\,\, |u_{\downarrow}^0| ^2 =q^2 |u_{\uparrow} ^0| ^2   .
\end{equation}
\par
The sum of these two terms equals the total electron density $n_{e0}$ at infinity,
$  |u_{\uparrow} ^0 | ^2  + |u_{\downarrow} ^0 | ^2  =n_{e0}\,\, .$
\par
We introduce the dimensionless function $f(x)$ as
$u_{\uparrow} (x)={u_{\uparrow} ^0 }f(x)$
and Eq. (\ref{glh5xym}) has the form
\begin{equation}\label{glh10hm}
-\xi _{\vec H} {^2}(T)\frac{d^2 f}{dx^2}-f+f^3  =0\,\, ,\,\,\,
 \xi _{\vec H}(T)=\sqrt{\frac{\hbar ^2}{2m|\tilde{\alpha}|}}
\end{equation}
with the solution
$f(x)=\tanh\left[\frac{x}{\sqrt{2}\xi _{\vec H} (T)}\right]$.
$\xi _{\vec H}$ and $f(x)$ are the functions of the magnetic field $\vec H$.
\par
The next two matrix rows of Eq. (\ref{glpinh3nm}) can be examined in the same way. So, we get a solution of  Eq. (\ref{glpinh3nm}) in the explicit form
$\Upsilon ^{+} =(u_{\uparrow}\, ,\, u_{\downarrow}\, ,\, v_{\uparrow}\, ,\, v_{\downarrow})$.
The expression
enables to analyze
 the surface tension due to the order parameter $\Upsilon$ in the magnetic field $\vec H$.
The free energy $\emph{F}_{\Upsilon ,\vec H}$ due to the order parameter $\Upsilon$ and $\vec H$ per unit area of the interface is
\begin{eqnarray}\label{f100}
\emph{F}_{\Upsilon ,\vec H}=\int_{0}^{\infty}dx\, \digamma _{\Upsilon,\vec H}\equiv
\int_{0}^{\infty}dx\emph{F}_{\Upsilon}\Upsilon (\vec r))+ \nonumber
\\
\int_{0}^{\infty}dx\left[
\emph{F}_{para}(\vec {H}+
4\pi\vec{M} ,\Upsilon , \Upsilon ^{+})-\vec H\vec M -\frac{{\vec H}^2}{8\pi}
\right]
\end{eqnarray}
The surface energy $\Sigma _{\Upsilon , \vec H}$ represents the difference between
$\emph{F}_{\Upsilon ,\vec H}$ and the free energy
of the uniform metal phase filling the half-space of the specimen. In the paramagnetic case
(\ref{f100}) with $\vec M(\vec r)\equiv 0$, the two last terms in the integrand
do not give contribution into $\Sigma _{\Upsilon , \vec H}$.
The calculation  is straightforward (cf.\cite{dG}) and gives $\Sigma _{\Upsilon ,\vec H}$:
\begin{equation}\label{gln20x}
\Sigma _{\Upsilon ,\vec H}=\frac{2\sqrt{2}}{3}\xi _{\vec H} (T)\tilde{\beta} n_0 ^2
\end{equation}
Here $n_0$ is the bulk density of electrons in metal. And $n_0 =\left({\Upsilon ^{+}
\Upsilon }\right)|_{|\vec r|\rightarrow \infty}$ , see (\ref{glr2}).
If we put $n_0$ to be of the order of an unit per
crystal cell we get for  $\Sigma _{\Upsilon , \vec H}$  (\ref{gln20x}) the typical value for metals. In addition,
 $\Sigma _{\Upsilon, \vec H}$ is a singular
function in the limit $ n_{0}\Rightarrow 0$, namely $\Sigma _{\Upsilon ,\vec H}\sim n_{0}^{3/2}$ as
$ n_{0}\Rightarrow 0$. And also $\Sigma _{\Upsilon ,\vec H}\sim (\rho -\rho _{c \vec H})^{3/2}$ in the limit
$ \rho\Rightarrow \rho _{c \vec H},\,\,\,\,\,\, \rho _{c \vec H}=\rho _{c 0}-\mu _{ph} Hsgn{H_x}a^{-1}$ .
\par
Now we consider the second situation
\begin{equation}\nonumber 
-\frac{\hbar ^2}{2m}\left(\frac{d^2 \Upsilon}{dx^2}\right)
+\alpha\Upsilon+\beta \left({\Upsilon ^{+}\Upsilon}\right) \Upsilon -\vec H\frac{\partial\vec M}{\partial ( \Upsilon ^{+}\Upsilon )}\Upsilon  =0
\end{equation}
We assume that $\vec M(\vec r)$  depends on electron density $n(\vec r)$ or
$\left({\Upsilon ^{+}\Upsilon }\right)$  and does not depend on $\rho (\vec r)$, i.e.
\begin{equation}\nonumber
\vec M(\vec r)=\vec M(n(\vec r))\,\, ,\,\,\, n(\vec r)=\Upsilon ^{+} (\vec r)\Upsilon (\vec r)
\end{equation}
In this case we can put components of
$\Upsilon (x)$ and $\Upsilon ^{+}(x)$ in the form of real functions  which do not vary in spin
space and depend on $x$:
$\Upsilon ^{+} (x)=\chi (x)
( s_{u\uparrow}, s_ {u\downarrow }, s_{v\uparrow}, s_{v\downarrow})$,
$s_{u\uparrow}, s_ {u\downarrow }$ and $s_{v\uparrow}, s_ {v\downarrow }$ are components of
two vectors in
spin space and are normalized by the condition  $$\Upsilon ^{+}(x)\Upsilon (x)=\chi ^2 (x)=n(x)\, . $$
 We arrive at the equation  for the scalar
 function $\chi (x)$:
\begin{equation}\label{Ap5}
\frac{\hbar ^2}{2m}\frac{d^2 \chi}{dx^2}=\alpha \chi +\beta \chi ^3 -\vec H\frac{\partial\vec M}{\partial (n(x))}\chi
\end{equation}
If we assume $\lim |_{\chi \Rightarrow 0}\,\,\,
\left\{\chi *\frac{\partial\vec M}{\partial (n(x)}\right\}\Rightarrow 0 $,
the Eq.(\ref{Ap5}) has two uniform solutions: 1) the  solution $\chi \equiv 0$ for the insulating phase,
2) the solution $\chi _{0}$ for the metallic phase:
 $\chi _{0} ^2 =|\alpha |/\beta \, .$
The function $\chi(x)$
satisfies the following boundary conditions:
$ \chi =0 ,\,$ at $\, x=0$ and
$ \chi\Rightarrow \chi _{0}\, ,\,\,$ at $ x\Rightarrow\infty$.
We multiply (\ref{Ap5}) by ${d\chi}/{dx}$ and find the first integral of it:
\begin{equation}\label{Ap6}
-\frac{\hbar ^2}{2m}\left(\frac{d \chi}{dx}\right)^2-|\alpha|\chi ^2 +
\frac{\beta}{2}\chi ^4 -\vec H\vec M (\chi ^2)=C_0
\end{equation}
$C_0$  can be found from the boundary conditions. So,
$$C_0=-|\alpha|\chi _{0}^2 +\frac{\beta}{2}\chi _0 ^4 -\vec H\vec M (\chi _{0}^2)
\, ;\,\,\chi ^2 (x)\equiv n(x)\, .$$
\begin{eqnarray}\nonumber
\frac{\hbar ^2}{2m}\left(\frac{d \chi}{dx}\right)^2=-|\alpha|(\chi ^2 -\chi _{0} ^2)
 +\frac{\beta}{2}(\chi ^4
-\chi _{0} ^4)- \nonumber
\\
 -\vec H (\vec M (\chi ^2)-\vec M (\chi _{0}^2)).
\end{eqnarray}
From the last equation the function $\chi (x)$ can be extracted in the implicit form.
And the surface tension due to the order parameter $\chi$ can be written as follows:
\begin{eqnarray}\label{Ap11}
\Sigma _{\Upsilon}=\hbar\sqrt{\frac{\beta }{m}}\int_{0}^{\chi _0}d\chi\Gamma ^{1/2}(\chi )\theta (\Gamma (\chi ))\,\, ,\,\,\,\, \nonumber
\\
 \Gamma (\chi)=\left[(\chi _0 ^2 -\chi ^2 )^2+
2\beta ^{-1}\vec H (\vec M (\chi _{0}^2)-\vec M (\chi ^2))\right]
\end{eqnarray}
Here $\theta (\upsilon )=0\,\,$
for $\upsilon <0\,\,$ ;
$\,\,\theta (\upsilon )=1\,\,$ for $\,\,\upsilon >0\, .$
The presence of $\theta (\Gamma (\chi ))$ in the integrand of (\ref{Ap11}) leads to the appearance
of a singularity of the value $\Sigma _{\Upsilon}$ for $H=0$. It leads also to singularities in the points where $\vec M $ becomes zero.
If $\vec H\equiv 0$,
\begin{equation}\label{Ap11H}
\Sigma _{\Upsilon 0}=\Sigma _{\Upsilon}|_{H\Rightarrow 0}=\frac{2\hbar}{3\beta\sqrt{m}}|\alpha|^{3/2}
\end{equation}
Below we consider $\Sigma _{\Upsilon}$ for Pauli paramagnetism  and Landau diamagnetism
 \cite{LandLif}. In both cases the magnetic
moment $\vec {M} $ can be represented as
$\vec M=\mu _{H} \vec B$. In the case of the Pauli paramagnetism $\mu _{H}=\mu _{para} =1/2 *\left(1/2\pi\right)^2
\left(e^2/\hbar c\right)\left(p_F /mc\right)$. In the case of the Landau diamagnetism $\mu _{H}=\mu _{dia}=-\frac{1}{3}\mu _{para}$; $ p_F =(3\pi ^2)^{1/3}\hbar n^{1/3}$  is Fermi momentum.
$\Sigma _{\Upsilon}$ (\ref{Ap11}) equals:
\begin{eqnarray}\label{Ap11m}
\Sigma _{\Upsilon}=1.5*\Sigma _{\Upsilon 0}\int_{0}^{1}dx\Gamma _{1} ^{1/2}(x)\theta (\Gamma _{1}(x))\,\, ,\nonumber
\\
  \Gamma _{1}(x)=\left[
 (1 -x^2 )^2\pm D G_{\pm}(1- x^{2/3})\right]^{1/2}
\end{eqnarray}
$D=2\beta\mu _{eff}H^2/|\alpha | ,\,  \mu _{eff}=1/(2\pi)^2)(e^2/\hbar c)(p_{eff}/mc) , $
$$p_{eff}=\hbar \left(3\pi ^2\aleph \rho _c/\beta\right)^{1/3}\, .$$
Here $\mu _{eff}$ is the effective susceptibility and $p_{eff}$ is the effective Fermi momentum.
 In the low magnetic field ($D\ll 1$) we represent $\Sigma _{\Upsilon}$ as follows:
\begin{equation}\label{Ap11H1}
\Sigma _{\Upsilon}=\Sigma _{\Upsilon 0}+\Delta\Sigma _{\Upsilon}\, .
\end{equation}
In the cases of Pauli paramagnetism and Landau diamagnetism we have
\begin{eqnarray}\label{Ap11H2}
\Delta\Sigma _{\Upsilon , Pauli}\cong0.536 H^2 \frac{3^{1/3}}{4\pi ^{2/3}}\frac{\hbar e^2}{cm^{3/2}}\frac{1}{|\alpha|^{1/6}\beta ^{1/3}}\nonumber
\\
\Delta\Sigma _{\Upsilon , Landau}\cong -0.290 H^2 \frac{3^{1/3}}{4\pi ^{2/3}}\frac{\hbar e^2}{cm^{3/2}}\frac{1}{|\alpha|^{1/6}\beta ^{1/3}}
\end{eqnarray}
\par
When $T$ goes to $T_c$,  $|\alpha|\Rightarrow 0$ and $|\Delta\Sigma _{\Upsilon}|$ rises. The value is restricted by the condition  $D\ll 1$. It means a reduction of surface tension $\Sigma _{\Upsilon}$ in the case of Landau diamagnetism and an increase of it in the case of Pauli paramagnetism. The asymmetry of behavior of the value of $\Sigma _{\Upsilon}$ between Landau diamagnetism and Pauli paramagnetism increases for larger values of $D$. In the limit $D\gg 1$ we have for Pauli paramagnetism $\Sigma _{\Upsilon}\gg\Sigma _{\Upsilon 0}$:  
\begin{equation}\label{Ap11H3}
\Sigma _{\Upsilon ; Pauli}\cong G_{+} ^{1/2}|\alpha |^{-1}\mu _{eff} ^{1/2}H \Sigma _{\Upsilon 0}\sim |\alpha |^{2/3}\,\, ,  
\end{equation}
For Landau diamagnetism:
\begin{equation}\label{Ap11H3L} 
\Sigma _{\Upsilon ; Landau}\equiv 0\, ,\,\,\,\, D\succeq 4.645\,\, .\,\,\,\,\,\,\,\,\,
\end{equation}
It is seen from (\ref{Ap11m}) that $\Sigma _{\Upsilon}\geq 0$. It should be
emphasized that the value of surface tension in magnetic field $\Sigma _{\Upsilon}$ can strongly differ from the value of $\Sigma _{\Upsilon 0}$ in the absence of magnetic field. In particularly, $\Sigma _{\Upsilon}$ decreases for the case of Landau diamagnetism and can be done even zero. In the opposite case of Pauli paramagnetism $\Sigma _{\Upsilon}$ can be done substantially more than $\Sigma _{\Upsilon 0}$. This effect leads to a strong asymmetry of the phase transition kinetics for diamagnetic and paramagnetic materials in the magnetic field. This effect is very important even in low magnetic fields. It is connected with the exponential dependence of probability for thermal nucleation as a function of the value of surface tension. The probability in the
process of phase transition\cite{LandLif} equals approximately:
\begin{equation}\label{sigma}
W=\omega_{th}\exp{\left(-\frac{U_0}{T}\right)}s^{-1}\, ;\,\,\,\,\,\omega_{th}=\frac{\omega_0 V}{4\pi R_c ^3 /3}\, .
\end{equation}
Here $V$ is the volume of the system, $R_c$ is the critical radius of thermal fluctuations and $\omega_0$ is the frequency of heterophase fluctuations in the metastable phase, i.e. the characteristic frequency of small oscillations in the potential well of a nucleus of the stable phase. Usually, $\omega_0$ can be estimated numerically by the order of the magnitude as about the Debye frequency $\omega _D$. It should be emphasized that the value of $\omega_{th}$ is really great (cf.\cite{LifKagan, budu2}):
\begin{equation}\label{sigma1}
\omega_{th}\simeq\exp{( N_0)}\, s^{-1}\, , \,\,\, N_0\gg 1\, .
\end{equation}
For the system with the volume $V\approx 1 cm^3\, , \,\,\,\omega _D\approx 10^{11}s^{-1}$, and
the critical radius $R_c$ of about several interatomic distances, we have $N_0\approx 80$. We should note
that in real experiment the critical $R_c$ should be not more than several interatomic distances.
In opposite case (see e.g.\cite{budu2}), the time of decay of the metastable system ($\sim 1/W$)
 will exceed  the time of existence of our Universe.
The same dramatic situation takes place also for quantum nucleation\cite{LifKagan, budu2}.
Small corrections of $R_c$ of the order of $N_0 ^{-1}$ lead to modulation of the value $W$ of the order of unity (see (\ref{sigma1})). For example, if a variation of the surface tension is of the order ten percent, the modification of  $W$ is about $\sim 2500$ or $\sim 1/2500$. So, if $D\sim 0.1$ in Eq (\ref{Ap11m}), $W$ increases in $\sim 2500$ times for the case of Landau diamagnetism and decreases in $\sim 2500$ times for the case of Pauli paramagnetism. In the vicinity of phase transition the value $D\sim 1$ can be reached even for extremely small magnetic fields because $|\alpha|\Rightarrow 0$. So,  the probability for thermal nucleation $W$ in the presence of magnetic field for diamagnetic substances is much more than that for paramagnetic substances. Probably, this fact accounts for the well known diamagnetic behavior of all semimetals \cite{kik}.  The possible reason of this behavior is the kinetic nature of stable phase nucleation in the
metastable environment with a small barrier due to the small surface tension. So, the growth of diamagnetic semimetals is much more probable than of paramagnetic ones. To verify this assumption experimentally it is necessary to investigate the phase transition kinetics by shielding the magnetic field. The other possible verification can be done by carrying out the experiment in an altering in time magnetic field.
\par
In summary, it is built a new approach to MI phase transition kinetics. It is based on two order parameters of sufficiently different nature. The equations of GL type are formulated for these two order parameters. Effects of magnetic field are included in the kinetics of MI phase transition kinetics. The exact solution of equations is obtained for paramagnetic case
(\ref{glpinh3nm}) with nontrivial behavior as a function of the uniform magnetic field $\vec{H}$. It is shown a radically different behavior of MI phase transition for Landau diamagnetism and Pauli paramagnetism cases.
\par
The author thanks S.N.Burmistrov and Yu.Kagan for valuable critical comments.
The work is supported in part by the Grant of RFBR No 16-02-00382.


\begin{thebibliography}{99}
\bibitem{NFMott}  N.~F.~Mott, "Metal--Insulator" Second Edition Taylor@Francis London--New York--Philadelphia,1990.
\bibitem{LandLif} L.D. Landau, E.M. Lifshitz . Statistical Physics. Part 1: Course of Theoretical
Physics, Vol. 5 (3rd ed.) Publisher: Butterworth-Heinemann; 3 edition, 1980.
\bibitem{Lubensky} P.~M.~Chaikin and T.~C.~Lubensky, Principles of Condensed Matter Physics,
Cambridge university press, 1995.
\bibitem{JinwuYe} Jinwu Ye, Phys. Rev. Lett., \textbf{97}, 125302 (2006)
\bibitem{dub} L.~Dubovskii,  JETP Letters \textbf{99}, No. 1, pp. 22-26 (2014).
\bibitem{dub1} L.~Dubovskii,    J. Low Temp. Phys. \textbf{182}, 192-205 (2016).
\bibitem{CahnHillard} J.~W.~Cahn and J.~E.~Hillard, J. Chem. Phys. \textbf{28} ,258 (1958);
 ibid. \textbf{42} , 93, (1964).
\bibitem{LifKagan} I.~M.~Lifshits and Yu.~Kagan. Sov. Phys. JETP 35, 206 (1972)
\bibitem{LandLif2} L.~P.~Pitaevskii, E.~M.~Lifshitz. Statistical Physics, Part 2. Vol. 9 (1st ed.), 1980.
\bibitem{kik} Tables of physical quantities. Editor I. K. Kikoin. M., Atomizdat, 1976 (in Russian).
\bibitem{Partenskiy} M.B. Partensky UFN, 128, 69 (1979).
\bibitem{VRG} G.~Vignale, M.~Rasolt, D.~J.~W.~Geldart, Adv. Quantum Chem. \textbf{21}, 235 (1990).
\bibitem{VR} G.~Vignale and M.~Rasolt, Phys. Rev. Lett. \textbf{59}, 2360 (1987); G.~Vignale,
PRB \textbf{37} 10865 (1988).
\bibitem{GL} V.L. Ginzburg and L.D. Landau, Zh. Eksp. Teor. Fiz. 20, 1064 (1950).
English translation in: L. D. Landau, Collected papers (Oxford: Pergamon Press, 1965) p. 546
\bibitem{dG} ~P.~G.~De~Gennes. Superconductivity of Metals and
Alloys, ~W.~A.~Benjamin, INC. New York, 1966.
\bibitem{MEG} E.~G.~Maksimov.
Phys. Usp. 53, 1185–1190 (2010).\bibitem{RC} A.~K.~Rajagopal and J.~Callaway, Phys. Rev. \textbf{B7}, 1912 (1973).
\bibitem{GH} C.~J.~ Grayce, R.~A.~Harris, Phys. Rev. \textbf{A50}, 3089 (1994).
\bibitem{LLe} L.D. Landau, E.M. Lifshitz.  Vol. 8. Electrodynamics of Continuous Media
(2ed., Pergamon), 1984.
\bibitem{AAA} A.A. Abrikosov.  Fundamentals of the theory of metals,
NY Butterworth-Heinemann, 1988.
\bibitem{budu2} S.~N.~Burmistrov and L.~B.~Dubovskii, Sov. Phys. JETP 66, 414 (1987).
\end{thebibliography}
\end{document}